# Reversible 1:1 Inclusion Complexes of $C_{60}$ Derivatives in α- and β-Cyclodextrins: Implications for Molecular Recognition-Based Sensing and Supramolecular Assembly

Guang-Zhong Yin, De-Yi Wang *

*IMDEA Materials Institute, C/Eric Kandel, 2, 28906 Getafe, Madrid, Spain*

**Corresponding Author**

*Tel: +34 91 549 34 22, fax: +34 91 550 30 47; Email: deyi.wang@imdea.org

**Abstract**

We developed a novel pathway to highly efficiently synthesize the [60]Fullerene ($C_{60}$)-Cyclodextrin (CD) complex, which is termed as solvent-induced reversible recognition. Three kinds of typical complexes, namely, $C_{60}$-2GUA$^+$-α-CD, $C_{60}$-2GUA$^+$-β-CD and $C_{60}$-4OH-β-CD, were synthesized. The chemical structure of all complexes was fully confirmed by Fourier transform infrared spectroscopy, Nuclear magnetic resonance (NMR), UV-Vis spectroscopy and the solubility test. Furthermore, crystalline status and thermal degradation were recorded by X-ray diffraction and Thermogravimetric analysis, respectively. Both the NMR and ESI-MS results clearly confirmed the 1:1 complex of $C_{60}$ derivatives with CD. The pathway overcame the preparation limitation of traditional complex, and was extended to the prepare α-CD containing complex. To the best of our knowledge, it is the first time to report α-CD related complex. The geometric relation between $C_{60}$ and CD for all the three kinds of complexes were analyzed. We can achieve high efficiency to obtain solid-state complex products, which lays the foundation for the widely practical application of complex and provides enough experimental basis for the advanced application of $C_{60}$ and CD in supramolecular assembly. Notably, the solvent-induced reversibility and ionic instability of the complex provides a new possibility for the molecular recognition-based sensing.

**Key words:** Cyclodextrin; [60]Fullerene; Complex; Supramolecular assembly; Host-guest recognition

**ToC**

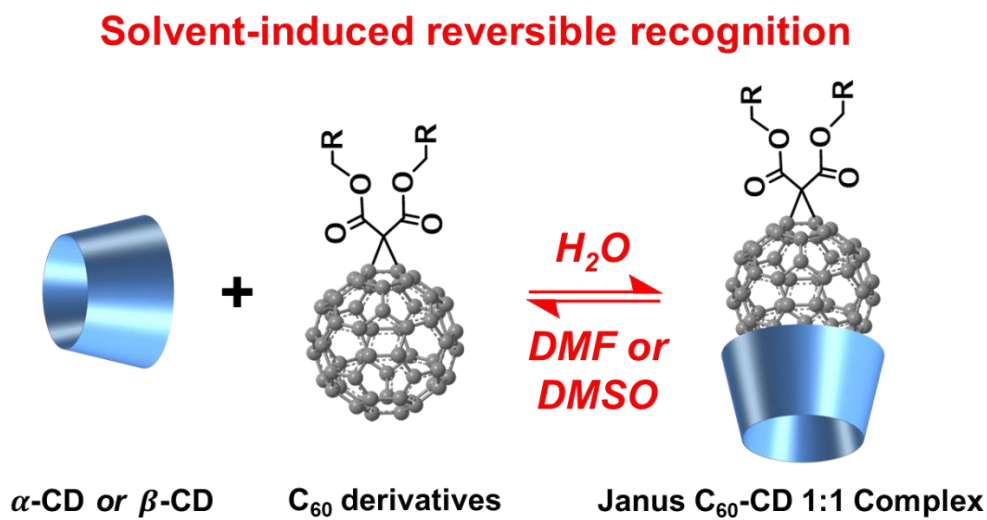

We developed a novel pathway to highly efficiently synthesize the [60]Fullerene ($C_{60}$)-Cyclodextrin (CD) complex, which is termed as solvent-induced reversible recognition.

**HIGHLIGHTS**

1. Solvent-induced reversible recognition was developed as an efficient way to prepare CD-$C_{60}$ complex. We can prepare solid-state products with high conversion and high efficiency by this method.

2. It is the first time to report the *α*-CD containing complex. It is determined that the signal at near 1510 $cm^{-1}$ in FTIR is the characteristic vibration absorption of CD-$C_{60}$ complex.

3. The relative spatial structure of complex is analyzed by using simple geometry. The size of three classical CDs was analyzed systematically.

4. The solvent-induced reversibility and ionic instability the complex provides a new platform for supramolecular assembly and molecular recognition-based sensing.

# 1. Introduction

[60]Fullerenes ($C_{60}$) and their derivatives have attracted a lot of attention in recent years and have been successfully applied to materials science and biological technology.[1-2] However, the practical use of these potential biomedical applications of $C_{60}$ have been hampered, owing to their low solubility in water and other common solvents.[3] Several approaches have been explored for preparing water-soluble fullerenes,[4] e.g. carboxylic acid fullerene derivatives,[5-6] and fullerol.[7-8] On the other hand, as multifunctional molecular receptors, cyclodextrins (CD) can selectively bind a wide variety of guest molecules through hydrophobic interactions, to form host-guest inclusion complexes or nanometer supramolecular assemblies.[9-11] Nowadays, investigations on the molecular recognition and assembly or self-assembly of CD and their derivatives are still one of current interest in chemical and biological systems. $C_{60}$-CD related complex is a kind of functional compound which can combine the advantages of CD structure and fullerene function. It has been widely studied in the past decade, and a series of complex have been synthesized.[12-15] Due to the large size of γ-CD and its high degree of fit with $C_{60}$, it is easy to form γ-CD-$C_{60}$ complex.[16-21] For example, Iohara *et al.*[22] reported hydrophilic $C_{60}$ nanoparticles that are highly stable in living systems were prepared with an anionic γ-cyclodextrin derivative, via a simple procedure for use in biological applications. Recent studies show that β-CD can also be widely used to prepare complex.[23] It is reported that the inner diameters of the β- and γ-CD cavity are only 0.78 nm and 0.95 nm, respectively, while the diameter of $C_{60}$ is estimated to be about 0.71 nm. Mainly because of this dimensional limitation, it is generally thought that complete inclusion of $C_{60}$ is not possible and a 2: 1 (CD: $C_{60}$) complex forms.[23-25] For a typical method to prepare CD-$C_{60}$ complex, it needs a large quantity of mixed solvents (with high boiling temperature), which are difficult to remove,[23] and the state of the product is mostly low concentration solution. These limitations greatly increase

the production cost and reduces the production efficiency, and directly limits the application of corresponding complex. In addition, the type of CD also seems to be limited because the cavity size of α-CD is too small. To the best of our knowledge, there are no reports about α-CD-$C_{60}$ complex at present.

As it is well reported, $C_{60}$-CD complex can be widely used in photodynamic therapy, self-assembly [26], size-exclusive nanosensors [27] and nanocomposites [28]. In consideration of its extensive potential advanced application value of $C_{60}$-CD complex and the limitations of current reported methods, it is of practical scientific significance to further developing new pathway, and exploring new complex structure as well as the new high-performance application of the complex. To develop an efficient pathway for preparing $C_{60}$-CD complex and to expand the application of α-CD in the preparation of complex, in this work, we will design and prepare some typical novel $C_{60}$-CD complexes, which contains α-CD or β-CD, and modified $C_{60}$ with neutral or ionic groups. The preparation method, chemical structure, solubility, aggregation structure and thermal decomposition behavior of the complexes will be considered sufficiently.

## 2. Structure analysis

**Scheme 1** Solvent-induced reversible recognition to synthesize serials of novel $C_{60}$-CD complexes.

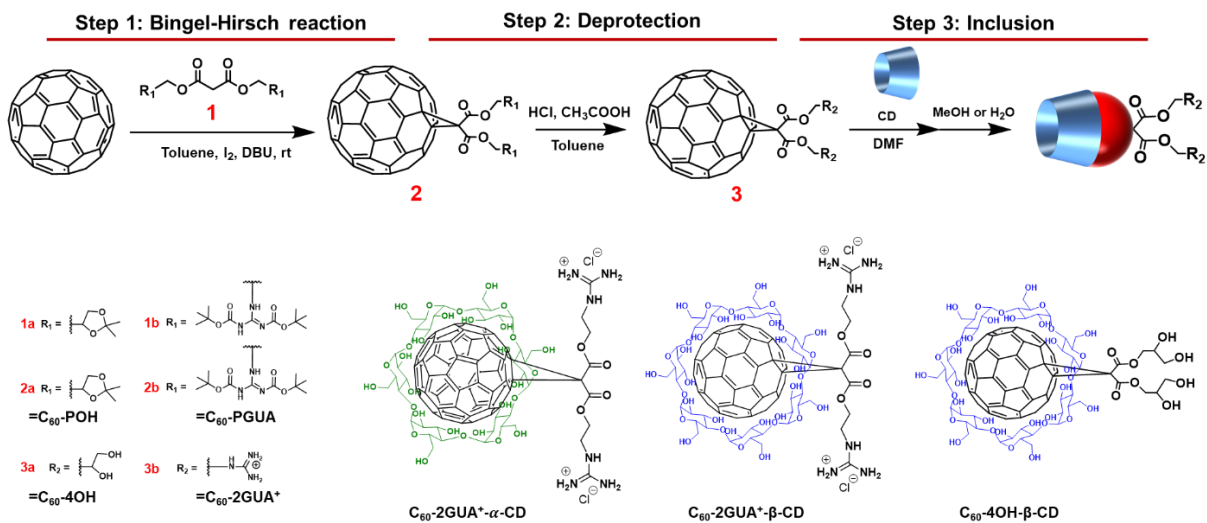

## 2.1 Synthesis of the complexes and chemical structure analysis

As shown in Scheme 1, the complex was synthesized following the three steps: (1) Bingel-Hirsch reaction to synthesize compound **2** (including **2a** $C_{60}$-POH and **2b** $C_{60}$-PGUA); (2) deprotection to generate novel $C_{60}$ derivatives (compound **3**) with polar functional groups (namely, **3a** $C_{60}$-4OH and **3b** $C_{60}$-2GUA$^+$) and (3) solvent induced reversible recognition to form the three typical complexes. Typically, $C_{60}$-PGUA and $C_{60}$-POH were synthesized according to Bingel-Hirsch reaction, which was widely reported in the literature elsewhere. [29-30] After that, the $C_{60}$-4OH and $C_{60}$-2GUA$^+$ were formed after deprotection by using HCl according to the report elsewhere. [31] The synthesis details were all provided in Supporting Information S1, Experimental section. The successful synthesis of the compounds was fully confirmed by the FTIR, $^1$H NMR, $^{13}$C NMR and ESI-MS (Figure S1-S23).

By dissolving a solution of CD ($α$-CD or $β$-CD) and fullerene derivative ($C_{60}$-4OH or $C_{60}$-2GUA$^+$) in DMF and then adding the water dropwise, we could achieve very high yields of complex (≥90 %) product as red-brown precipitation. It is worth noting that the main design principal is realizing that one side of $C_{60}$ remains unchanged and the other side is hydrophilic, giving rise to the water phase stability of complex molecule. At present, a large number of articles have reported the modified fullerene based complex. [32] However, there are few reports on efficient preparation methods based on $α$-CD and $β$-CD, especially $α$-CD. We also tried to prepare the complex based on compound 2a and 2b (Scheme 1 and Figure S24). It was found that the solubility of the two derivatives was quite different from that of cyclodextrin, so it was difficult to form a suitable concentration of compound 2a and 2b in DMF, which hindered the efficient formation of the final complex. Therefore, the key characteristic of fullerene derivatives in this work is that they have quite similar solubility to CD, that is to say they can dissolve well in both DMF to form co-solution with CD. Furthermore, based on some simulation reports [28, 33], we think that complex exists as the molecular structure illustration in Scheme 1. Notably, the product is formed almost at the same time when water is added, which shows that the process of the inclusion is instantaneous and efficient. We directly proved the successful synthesis of complex by ESI-MS test. The specific data are shown in Figure S25-S26 (For complex $C_{60}$-2GUA$^+$-$\alpha$-CD, ESI-MS: Calcd.: 1966.78 (M-HCl), Found: 1968.12 (M+H$^+$), and for complex $C_{60}$-4OH-$\beta$-CD, ESI-MS: Calcd.: 2105.86, Found: 2128.11 (M+Na$^+$) and 2143.14 (M+K$^+$)). We didn't directly detect the signal of $C_{60}$-2GUA$^+$-$\beta$-CD. However, we can prove the obtaining of $C_{60}$-2GUA$^+$-$\beta$-CD complex by other indirect means, e.g solubility and FTIR spectrum. We can speculate on the formation process of the complex: $C_{60}$ and CD are well dissolved in DMF; with the addition of highly polar solvent (e.g. $H_2O$ or $CH_3OH$), the increasing of polarity will make the dissolved $C_{60}$ derivatives tend to precipitate from the solvent; while in an appropriate polarity range, the cavity of CD retains the most suitable space, so the $C_{60}$ derivatives are generally close to the cavity of CD rather than self-crystallization and precipitation; with the further increase of polarity, complex formed steadily and precipitated rapidly due to the low solubility in the mixture solvent. Indeed, the solubility of the product is significantly different from free $C_{60}$ derivatives and CD (**Figure 1a**). As it can be found, CD (white powder) is soluble in water; $C_{60}$ derivatives are well soluble in carbon disulfide ($CS_2$). The product (red-brown powder with yielding of ≥90 %) is insoluble in $CS_2$ and difficult to be dissolved in water (in fact, it can be dissolved in water after long time and with a low concentration). This remarkable difference in solubility strongly suggests that the $C_{60}$ has been encapsulated, because $C_{60}$ derivatives dissolves mainly in non-polar or low-polar solvents such as $CS_2$, DMF or DMSO, while the encapsulation in the CD cavity can make $C_{60}$ derivatives soluble in water. To ensure that no free CD or free $C_{60}$ derivatives remained, all samples were washed with excess of $CS_2$ in which the $C_{60}$ derivatives is soluble and excess of water in which the CD is soluble, respectively.

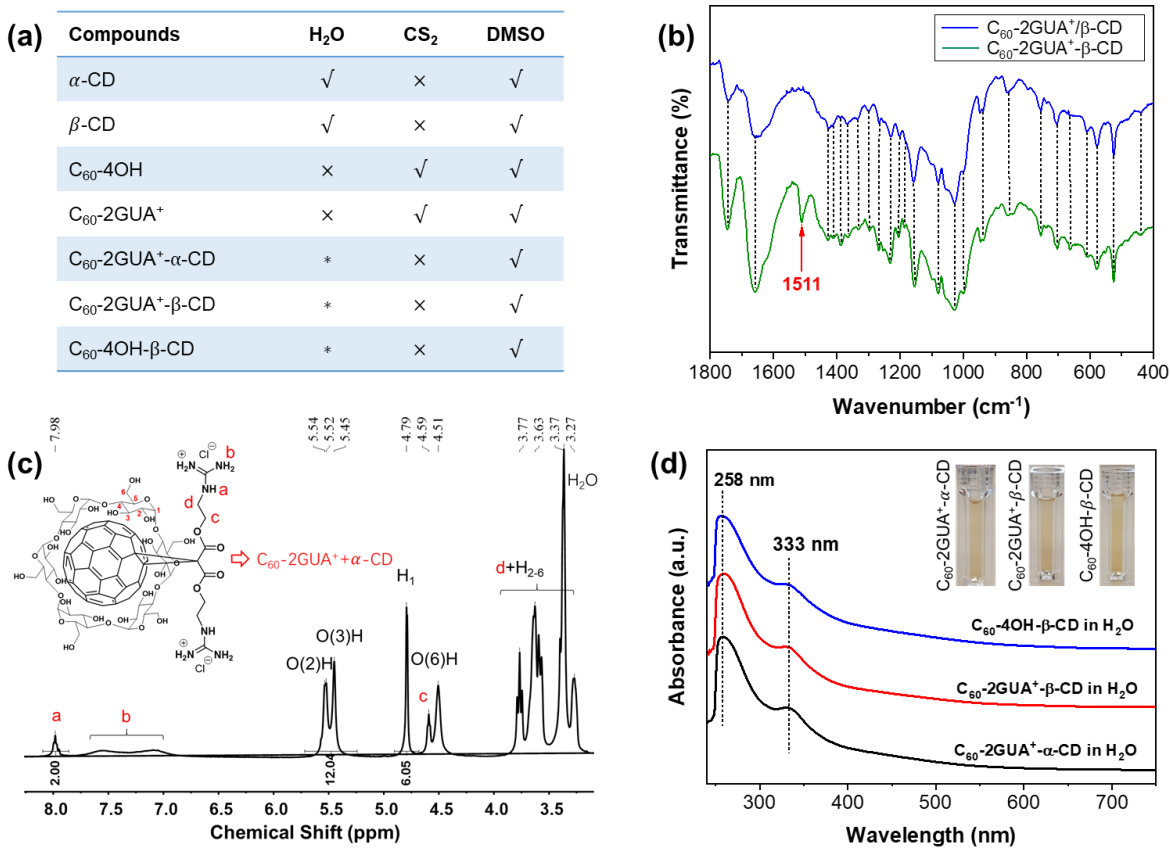

**Figure 1** (a) The solubility test results (Note: √ means the compound can be dissolved in the corresponding solvent; × means the compound cannot be dissolved in the corresponding solvent; * means the compound can be slightly dissolved in the corresponding solvent); (b) FTIR of signal comparing between $\beta$-CD/$C_{60}$-GUA$^+$ mixture and $C_{60}$-2GUA$^+$-$\beta$-CD complex; (c) $^1$H NMR spectrum obtained from sample of $C_{60}$-2GUA$^+$-$\alpha$-CD that dissolved in DMSO-d$_6$; and (d) UV-Vis spectra of $C_{60}$-2GUA$^+$-$\alpha$-CD, $C_{60}$-2GUA$^+$-$\beta$-CD and $C_{60}$-4OH-$\beta$-CD with concentration of ~0.1 mg/mL in water.

We will further verify the successful synthesis of the complexes in the following. The detailed

FTIR and the assignments were all provided in Figure S21-S23. Here we choose sample $C_{60}$-2GUA$^+$-$\beta$-CD as a typical example to analyze the useful structure information. As shown in **Figure 1b**, comparing $C_{60}$-GUA$^+$/$\beta$-CD mixture and $C_{60}$-2GUA$^+$-$\beta$-CD complex, the only significant difference is the signal at 1511 cm$^{-1}$, which is assigned to the characteristic vibration signal of the $C_{60}$-CD complex (Table S1). Obviously, the FTIR spectrum shows the characteristic absorption peaks of $C_{60}$ at 524 cm$^{-1}$.[34] Furthermore, in the FTIR of $C_{60}$-2GUA$^+$-$\alpha$-CD and $C_{60}$-4OH-$\beta$-CD, the same signals are also obviously detected (Figure S22 and S23). Typically, the characteristic vibration signal in $C_{60}$-2GUA$^+$-$\alpha$-CD and $C_{60}$-4OH-$\beta$-CD appears at 1513 cm$^{-1}$ and 1507 cm$^{-1}$, respectively. Notably, the signal at about 1510 cm$^{-1}$ was shown in some other reports [35-36]. Therefore, we know that the same signal can be also detected in other host guest recognition systems based on cyclodextrin, which indicates this is a universal phenomenon. Here it is assigned that we can use FTIR results to confirm the formation of complex by the detection of the moderately strong signal at about 1510 cm$^{-1}$. Furthermore, the peak at about 1510 cm$^{-1}$ in the FT-IR spectrum represents the C=C stretch which is normally appeared at 1581 cm$^{-1}$,[37-38] because the complexation, part of the $C_{60}$ is embedded in CD, thus the C=C stretch of part of $C_{60}$ is constrained, which is in accordance with some other supporting references.[39]

**Figure 1c** shows the $^1$H NMR spectra of $C_{60}$-2GUA$^+$-$\alpha$-CD with clear assignments. As it can be seen, the signal at 7.96 ppm is the hydrogen signal of -NH-. The broad signal at ~7.06 ppm (signal b) is the ionic hydrogen signal in terminal -NH$_2^+$.[40] The signal at 5.50 ppm and 5.43 ppm are O(2)H and O(3)H in $\alpha$-CD, respectively. Notably, its integral area ratio of signal **a** (7.98 ppm) and signal O(2)H + O(3)H is consistent with 2:12, which can directly and fully prove the formation of 1:1 complex in the precipitates. Similar integral area results of the additional two complexes and nuclear magnetic spectra ($^{13}$C-NMR) are all provided in Figure S6, S19 and S20). In fact, all the three complexes are unstable in DMSO, thus, the chemical shifts change between complex and the corresponding mixtures (see Table S2), and the NMR results will be further explained in *section 3.1*.

The UV-Vis spectra of the three complexes are shown in **Figure 1d**. Because it is difficult to accurately control the concentration of the samples, we just present the normalized data separately in order to make it clear. As it can be seen, the solution of the three complexes is light yellow in color (inset of Figure 1d) and show characteristic bands at 258 nm and 333 nm. It is reported that the broad absorption in 450-550 nm region is characteristic of the aggregated state of crystalline $C_{60}$ derivatives and is caused by close electronic interactions among adjacent $C_{60}$ molecules. [21] The absence of the broad absorption in 450-550 nm region suggests that $C_{60}$ does not form aggregates but well solubilized in the solvents. [41] Indeed, the inclusion interaction ensures the water solubility of the three complexes. The UV-Vis spectra of $C_{60}$ derivatives in DMSO are provided in Supporting information (Figure S27a and S27b). There is no significant difference between $C_{60}$ derivatives and the solution obtained by dissolving complex in DMSO. The UV-Vis absorption curves of $GUA^+$-$C_{60}$ and $C_{60}$-$GUA^+$-$\beta$-CD were also measured (Figure S27c and S27d). The signal shift or shape change indicates that, $\beta$-CD can significantly improve the dispersion of $C_{60}$-$GUA^+$ in water after the formation of complex.

## 2.2 Geometric structure analysis of complex molecules

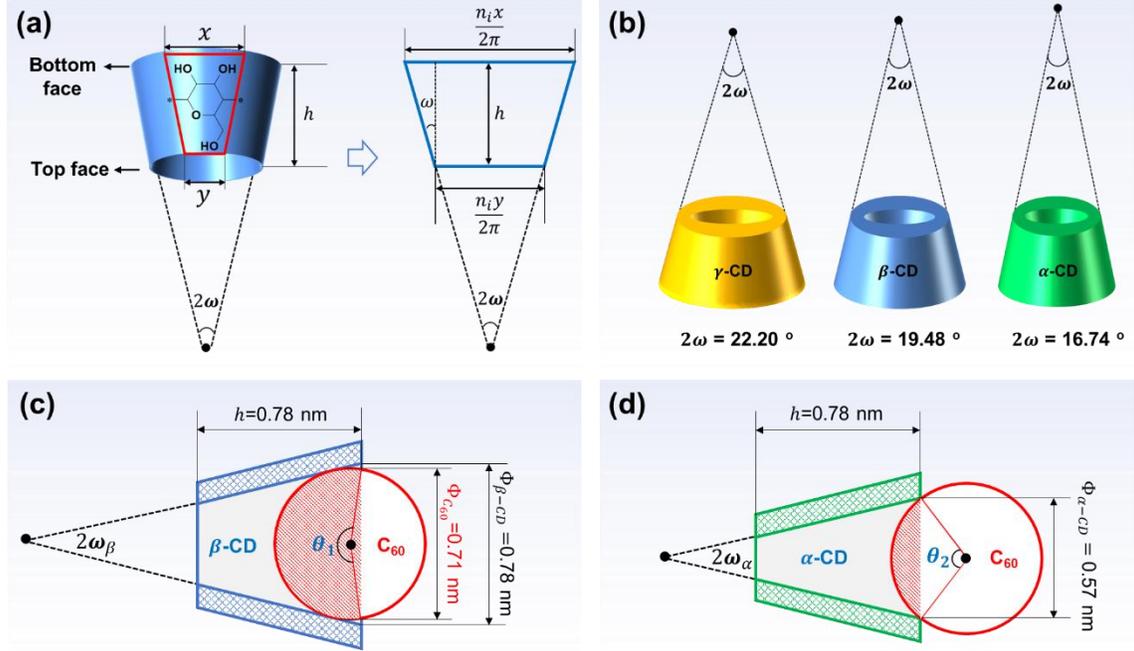

**Figure 2** (a) Taking $\beta$-CD as an example to abstract the geometric parameters of CD; (b) the numerical results of the cone angles of $\alpha$-CD, $\beta$-CD and $\gamma$-CD; (c) Geometric relations between $\beta$-CD and $C_{60}$ in complex and (d) Geometric relations between $\alpha$-CD and $C_{60}$ in complex.

Previous research further shows that $C_{60}$ is only accessible from the bottom face of CD.[41] To well understand the structure, we hope to interpret the complexes structure from the spatial level. We can obtain the detailed size of $\gamma$-CD according to the literature elsewhere,[42] as shown in Figure S28 and Table S3. Herein, we use some geometric derivation to obtain the necessary relevant dimensions for $\alpha$-CD and $\beta$-CD. The cross section of the CD can be abstracted as an isosceles trapezoid. The relevant dimensions and symbols are shown in **Figure 2a**. We assume the repeating elements have relatively fixed geometric parameters, that is to say the size $x$ and $y$ of the CD constituent elements (Equivalent arc length) are the same for all the $\alpha$-CD, $\beta$-CD and $\gamma$-CD, as shown in the red area in Figure 2a. So, we can get two diameter parameters of CD, namely, $n_i x/2\pi$ and $n_i y/2\pi$, which are the bottom and top of the isosceles trapezoid, respectively. Based on the model and equation (1) (See Figure S29 in the section S3.1 for the specific derivation process), and the specific parameters of $\alpha$-CD and $\beta$-CD reported in the literature and listed in Table S2,[43-44] we can calculate the cone angle of CD: $2\omega_{\alpha-CD}=16.74°$, $2\omega_{\beta-CD}=19.48°$, and $2\omega_{\gamma-CD}=22.20°$ (**Figure 2b**). The acquisition of the cone angle is very important for the geometric dimensions. For more detailed geometric dimensions of CD please see Figure S29 and Table S3.

$$\frac{tan\omega_{\alpha-CD}}{n_{\alpha-CD}} = \frac{tan\omega_{\beta-CD}}{n_{\beta-CD}} = \frac{tan\omega_{\gamma-CD}}{n_{\gamma-CD}} \quad (1)$$

where, $\omega_{\alpha-CD}$, $\omega_{\beta-CD}$, and $\omega_{\gamma-CD}$ are the half cone angle of $\alpha$-CD, $\beta$-CD and $\gamma$-CD, respectively; $n_{\alpha-CD}$, $n_{\beta-CD}$ and $n_{\gamma-CD}$ are the number of repeat unit of $\alpha$-CD, $\beta$-CD and $\gamma$-CD, respectively.

Based on the geometric relationship shown in **Figure 2c** ($\Phi_{\beta-CD} > \Phi_{C_{60}}$), the following relationship is derived:

$$\theta_1 = 360 - 2\arccos\left(\frac{\Phi_{\beta-CD}}{\Phi_{C_{60}}tan\omega_\beta} - \frac{1}{sin\omega_\beta}\right) \quad (2)$$

where, $\theta_1$ is the sphere cone angle of the overlap of C$_{60}$ and $\beta$-CD, $\Phi_{\beta-CD}$ is diameter of $\beta-CD$, and $\Phi_{C_{60}}$ is diameter of $C_{60}$ (0.71 nm [45-46]). Based on the geometric relationship shown in **Figure 2d** ($\Phi_{\alpha-CD} < \Phi_{C_{60}}$), the following relationship is derived:

$$\theta_2 = 2\arcsin(\Phi_{\alpha-CD}/\Phi_{C_{60}}) \quad (3)$$

where, $\theta_2$ is the sphere cone angle of the overlap of C$_{60}$ and $\alpha$-CD, $\Phi_{\alpha-CD}$ is diameter of $\alpha-CD$. The specific derivation process of eq. (2) and (3) is described in section S3.2 (Figure S30 and S31). Therefore, $V_{in,\ C_{60}-\alpha-CD}$ and $V_{in,\ C_{60}-\beta-CD}$, the inclusive volume ($V_{in}$, which is the volume of shadow part as shown in Figure 2c and 2d) can be further calculated via the equation (4) and (5), respectively:

$$V_{in,\ C_{60}-\alpha-CD} = \frac{2}{3}\pi R_{C_{60}}^3\left(1-\cos\frac{\theta_2}{2}\right) - \frac{1}{3}\pi R_{C_{60}}^3 \sin^2\frac{\theta_2}{2}\cos\frac{\theta_2}{2} \quad (4)$$

$$V_{in,\ C_{60}-\beta-CD} = \frac{2}{3}\pi R_{C_{60}}^3\left(1-\cos\frac{\theta_1}{2}\right) + \frac{1}{3}\pi R_{C_{60}}^3 \sin^2(180-\frac{\theta_1}{2})\cos(180-\frac{\theta_1}{2}) \quad (5)$$

where, $R_{C_{60}}$ is the radio of $C_{60}$.

Herein, the effect of functional groups on the shape of $C_{60}$ are ignored. In other words, the modification of $C_{60}$ does not affect the geometry of $C_{60}$ (sphere), which can be proved by the structural model in some other literatures.[47] Notably, the value obtained here is the maximum of the calculated value. This is because we calculated these indexes based on the intrinsic size of each molecule, not the van der Waals size. The corresponding values of $\theta_i$ and $V_{in}$ are all listed in **Table 1**. We only consider the type of CD and assume that functionalization of $C_{60}$ will not affect the shape of their intrinsic sphere geometry, so the values of $\theta_i$ and $V_{in}$ of $C_{60}$-2GUA$^+$-$\beta$ − CD and $C_{60}$-4OH-$\beta$ − CD are the same. Notably, based on the accuracy of geometry, molecular weight and molecular structure of these complexes, we classify them as Janus giant molecules, the typical molecular nanoparticles (see references [48-49] for more details on giant molecules). According to the calculation based on Figure S32 and S33, the volumes of the side functional groups on $C_{60}$ are too large relative to the chamber of CD. Therefore, it is only possible to form the structure as show in Figure 2c and 2d.

**Table 1** The geometric parameters of the three complexes.

| Complex | $\theta_i$ (°) | $V_{in}$ ($nm^3$) |
|---|---|---|
| $C_{60}$-2GUA$^+$-$\beta$ − CD, $C_{60}$-4OH-$\beta$ − CD | 182.22 | 0.77 |
| $C_{60}$-2GUA$^+$-$\alpha$ − CD | 106.80 | 0.16 |

## 2.3. Crystalline status analysis

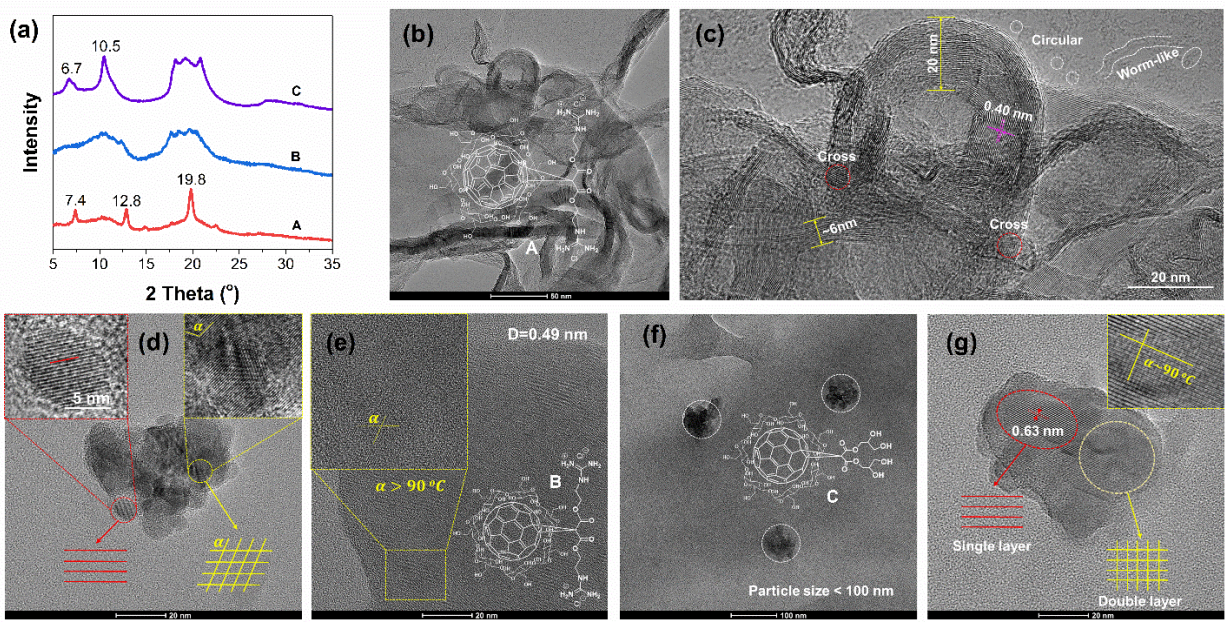

**Figure 3**. (a) XRD curves of A, $C_{60}$-2GUA$^+$-$\alpha$-CD, B, $C_{60}$-2GUA$^+$-$\beta$-CD, and C, $C_{60}$-4OH-$\beta$-CD, (b) Transmission electron microscope (TEM) image of complex A, (c) enlarged image based on Figure 3b, (d) TEM of complex B, (e) more details about the micro morphologies, (f) TEM of complex C and (g) enlarged image of complex C with clear crystal lattice.

The crystalline information of the CD, $C_{60}$ derivatives and complexes came from powder X-ray diffraction (XRD) measurements as shown in Figure 3(a). For neat $\alpha$-CD and $\beta$-CD, several salient peaks at $2\theta$ = 10.8º, 13.4º, 16.1º, 18.6º, 20.2º and 21.8º are observed in its XRD pattern, indicating a high degree of crystallinity (Figure S34-S36). XRD patterns of the three complexes in **Figure 3a** are very different from those of neat CD and $C_{60}$ derivatives. Characteristic peaks of CD (such as $2\theta$ =11.9º, 14.3º and 21.6º for $\alpha$-CD, and $2\theta$ =9.0º, 10.7º and 12.5º for $\beta$-CD) and $C_{60}$ derivatives (such as $2\theta$ =31.6º (Figure S37), $2\theta$ =10.3º for $C_{60}$-2GUA$^+$ and $2\theta$ =11.4º for $C_{60}$-4OH) were observed in none of the complexes, demonstrating the absence of free CD and $C_{60}$ derivatives in the complex samples. Due to the high symmetry of the original $C_{60}$ and CD, both show excellent crystallinity. After functionalization, the introduced side groups broke the symmetry of the molecules in varying degrees (**Figure 3b**), resulting in a significant decrease in the crystallinity of the $C_{60}$ derivatives under a certain crystallization condition.

The ordered belts (with size of 6-20 nm) and rings (Figure 3b-3c) and worm like supramolecular assemblies coexist simultaneously. The dark phase spacing of the strip is calculated about 0.40 nm. It is worth pointing out that, the cross angle is much high than 90 ° ($\angle\alpha$ in Figure 3d-3e) and the band spacing is 0.49 nm for the $C_{60}$-2GUA$^+$-β-CD. As shown in Figure 3f, $C_{60}$-4OH-β-CD sample presents aggregates about 100 nm in solution. (Figure 3f) The significant lattice stripes appear in the enlarged images. The spacing between the two lattice stripes is about 0.63 nm. The cross angle is almost 90° for the multilayer crystals, as shown in Figure 3g inset. Notably, the spacing of all lattice bands is less than the fullerene diameter (~ 0.71 nm). This is similar to the structure that reported in the literature.[50] This may be because error in the equipment measurement method. For complex containing the same fullerene derivative (sample $C_{60}$-2GUA$^+$-β-CD and $C_{60}$-2GUA$^+$-α-CD), there is a significant difference between the spacings of 0.49 and 0.40 nm. At the same time, both $C_{60}$-2GUA$^+$-β-CD and $C_{60}$-2GUA$^+$-α-CD are highly ordered, indicating that the type of cyclodextrin plays a key role in morphology, which can prove the formation of molecular complex to a certain extent. XRD results show that complex has a new crystalline structure. Through TEM characterization, there are indeed some nano aggregates in the solution. We obtained the average particle size by statistical processing of the images. Specifically, for example, for sample $C_{60}$-4OH-β-CD, nano crystals of 20-100 nm were formed. These microcrystalline particles correspond well with the wide signal peak in UV.

## 3 Stability of the complexes

## 3.1 Solvent stability: reversibility in solvent

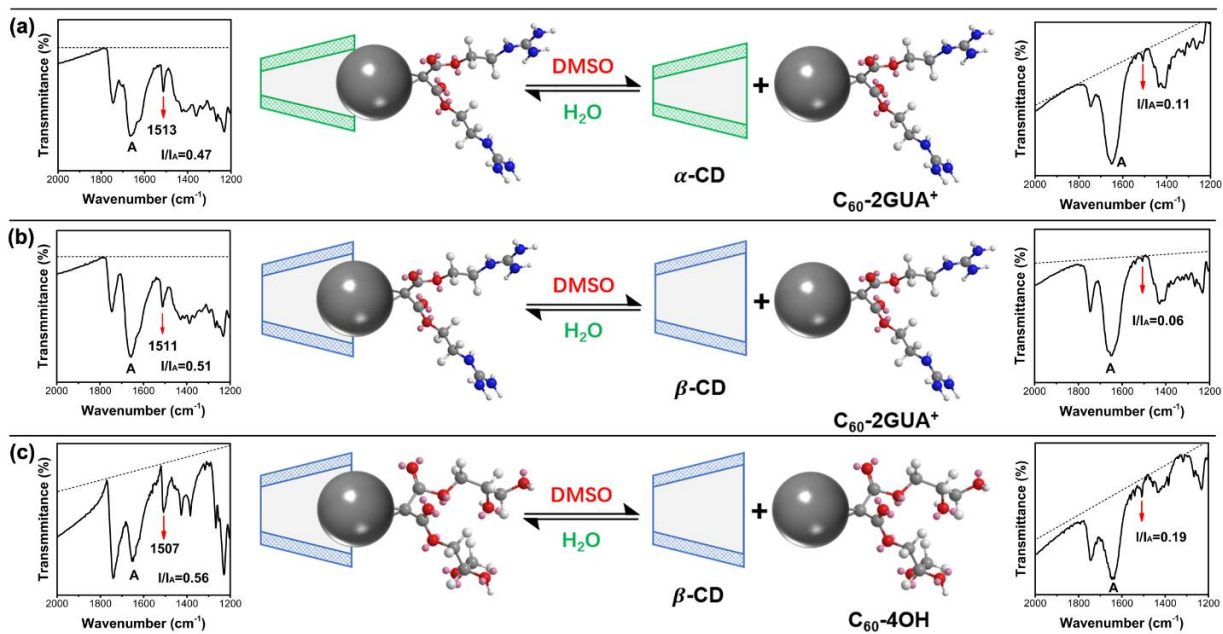

**Figure 4** The solvent induced reversibility illustration between the complexes and monomers (CD as well as $C_{60}$ derivatives): (a) $C_{60}$-2GUA$^+$-$\alpha$-CD, (b) $C_{60}$-2GUA$^+$-$\beta$-CD and (c) $C_{60}$-4OH-$\beta$-CD. FTIR was used to confirm the chemical structure. The left one is FTIR curve of a complex, and the right one is that of product from solution of the corresponding complex in DMSO (with concentration of 5 mg/mL).

It should be pointed out that we can use the characteristic signal near 1510 cm$^{-1}$ in FTIR to evaluate whether complex exists stably in the solvent by the relative signal strength. Typically, we choose DMSO as the solvent because DMSO is a good solvent for all compounds (Figure 1a). The DMSO was removed via volatilization at room temperature under gas flow, and then FTIR of the obtained solid was performed by KBr window method. **Figure 4** shows the solvent induced reversibility illustration between the complexes and monomers. The corresponding FTIR is also shown in Figure 4. As it can be seen, $I_A$ is the relative intensity of peak A (peak A is assigned to the signal from CD; Its strength is relatively constant (as shown in Figure S21-S23), so it is selected as the internal standard of FTIR analysis), $I/I_A$ is the intensity ratio of characteristic signal (~1510

cm$^{-1}$) and peak A, and the baseline is chosen according to the full FTIR curves (Figure S38). The characteristic peaks (near 1510 cm$^{-1}$) of the three complexes are significantly shown in the left three FTIR curves. After DMSO dissolution treatment, the characteristic peaks are significantly weakened or almost disappeared (the right three FTIR curves). As it can be seen, the $I/I_A$ for C$_{60}$-2GUA$^+$-$\alpha$-CD, C$_{60}$-2GUA$^+$-$\beta$-CD and C$_{60}$-4OH-$\beta$-CD are 0.50, 0.51 and 0.56, respectively. While after DMSO treatment, they are decreased significantly (0.11, 0.06 and 0.19 for C$_{60}$-2GUA$^+$-$\alpha$-CD, C$_{60}$-2GUA$^+$-$\beta$-CD and C$_{60}$-4OH-$\beta$-CD, respectively), which indicates the complexes are not stable in DMSO. In Figure 4, the accuracy of the specific index of characteristic peak intensity change based on FTIR is not so high, because the selection of baseline has certain subjectivity. But we can clarify the difference between the changes of the same compound. The same samples remained comparable before and after solvent treatment, because the area change is quite significant. That is to say, they will be dissociated in DMSO, giving rise to the mixture solution again. Therefore, the NMR results (Figure 1c) reflect the signals of the mixture of C$_{60}$ and $\alpha$-CD after the dissolution of complex C$_{60}$-2GUA$^+$-$\alpha$-CD. In addition, we found that the addition of inorganic salts (e.g. potassium bromide) to the aqueous solution of complex would rapidly precipitate to generate red solid (**Figure 5 a**). The FTIR results of the precipitate are shown in **Figure 5b**. The characteristic peak near 1510 cm$^{-1}$ disappeared completely, which indicated that complex could not exist stably under a certain ion concentration. Notably, the complexes have the immediate response of solvent and ion, which will provide a possible basis for the design of molecular recognition-based sensing in the future. The illustration of the mechanism of salt responsiveness can be shown in the following explanations (Figure 5c). The products formed by molecule-ion interactions are regarded as adducts instead of inclusion complex. [51]

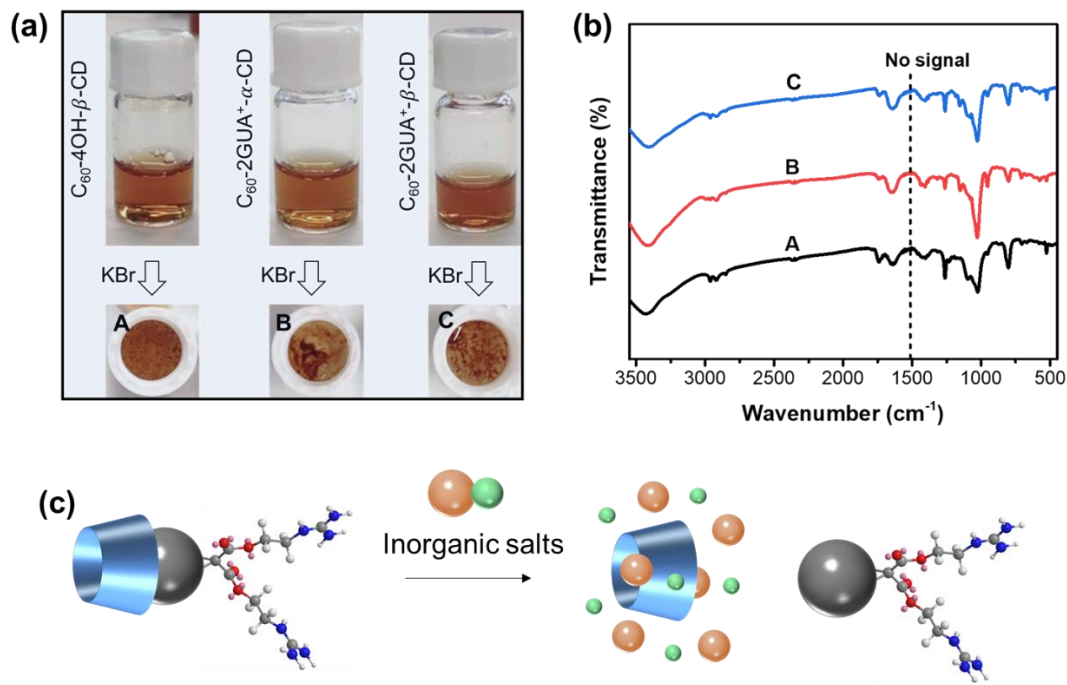

**Figure 5** (a) Image of aqueous solution (~2 mL with concentration of 0.5 mg/mL) of $C_{60}$-2GUA$^+$-$\alpha$-CD, $C_{60}$-2GUA$^+$-$\beta$-CD and $C_{60}$-4OH-$\beta$-CD and the corresponding precipitation phenomenon after adding some KBr (0.05 g), (b) FTIR curves of the precipitation (A is $C_{60}$-4OH mixing with $\beta$-CD; B and C are $C_{60}$-2GUA with $\alpha$-CD and $\beta$-CD, respectively.) in Figure 5(a) and (c) schematic diagram of possible principle of inorganic salt induced sedimentation.

## 3.2 Thermal stability

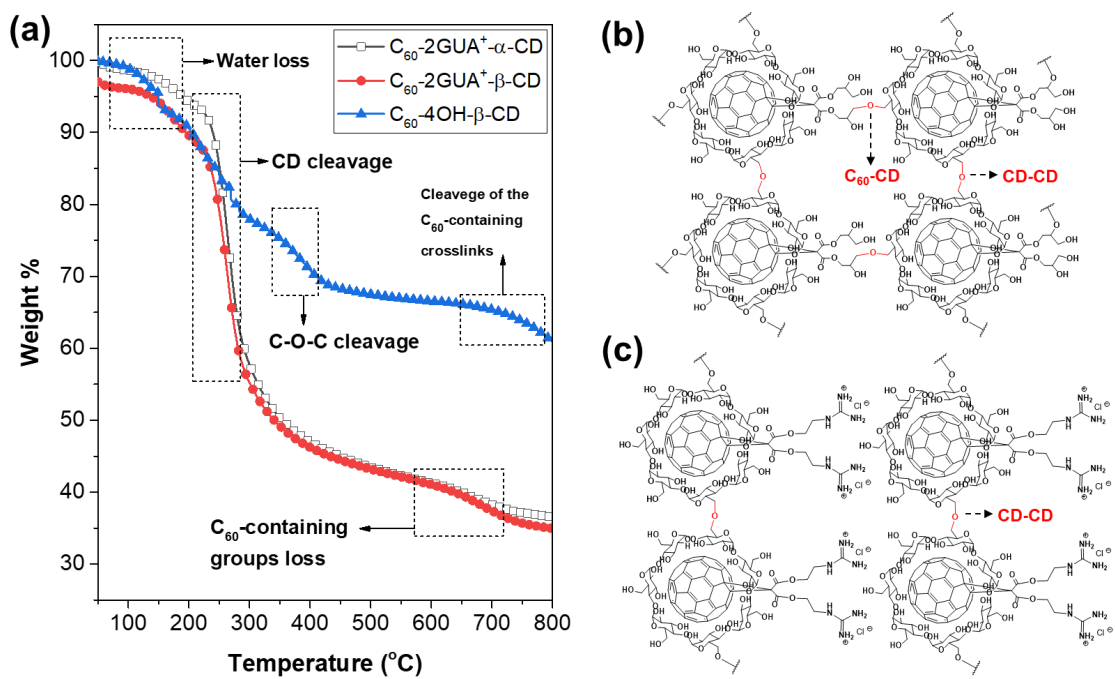

**Figure 6** (a) TGA curves of $C_{60}$-2GUA$^+$-$\alpha$-CD, $C_{60}$-2GUA$^+$-$\beta$-CD and $C_{60}$-4OH-$\beta$-CD, (b) The structure illustration of chemical crosslinking stabilization by intermolecular thermal dehydration (forming C-O-C between CD and CD (CD-CD), and between CD and $C_{60}$ (CD-$C_{60}$)) in $C_{60}$-4OH-$\beta$-CD, and (c) only C-O-C forming between CD and CD in $C_{60}$-2GUA$^+$-$\beta$-CD (The structure illustration of $C_{60}$-2GUA$^+$-$\alpha$-CD is the same as that of $C_{60}$-2GUA$^+$-$\beta$-CD)

The thermal decomposition behavior of the compounds was analyzed by TGA (Figure 6a and Figure S39). The neat $C_{60}$ has only one weight loss stage (600-700 °C, Figure S40). For $\alpha$-CD and $\beta$-CD, two weight loss stages are observed; the first weight loss (below 100 °C) is due to the evaporation of cavity water and the second one (~300 °C) corresponds to CD degradation (Figure S40). In contrast to the neat $C_{60}$ and CD degradation, the TG curves for all the complexes (**Figure 6a**) display different trend. The weight loss below 200 °C results from water release,[52] which is considered to be intermolecular dehydration condensation.[53] The weight loss of 250-350 °C is due to the cleavage of CD. The weight loss above 600 °C is the decomposition of $C_{60}$-containing fragments. Furthermore, from **Table 2**, we can see the Measured Re. (%) are 36.8%, 33.9% and 61.0 % for complex $C_{60}$-2GUA$^+$-$\alpha$-CD, $C_{60}$-2GUA$^+$-$\beta$-CD and $C_{60}$-4OH-$\beta$-CD, respectively. They were all much higher than corresponding Calculated Re. (%) (15.3 %, 15.8 % and 30.3 % for complex $C_{60}$-2GUA$^+$-$\alpha$-CD, $C_{60}$-2GUA$^+$-$\beta$-CD and $C_{60}$-4OH-$\beta$-CD, respectively). This demonstrates that complex formation improves the thermal stability of both CD and $C_{60}$ derivatives component, which can be attributed to the strong inclusion interactions between $C_{60}$ and CD. Specially for sample $C_{60}$-4OH-$\beta$-CD, the thermal degradation of 300-400 °C is due to the decomposition of some other organic groups (e.g. C-O-C, being generated via the crosslinking between $C_{60}$-4OH and CD) [54], which plays a main role in stabilizing compounds to a large extent (**Figure 6b**). As a comparison (**Figure 6c**), we know that there is no intermolecular dehydration between $C_{60}$-2GUA$^+$ and CD for both $C_{60}$-2GUA$^+$-$\alpha$-CD and $C_{60}$-2GUA$^+$-$\beta$-CD.

**Table 2** The thermogravimetric analysis of residue (at 800 °C)

| **Samples** | $\omega_{CD}$ | $\omega_{C_{60}}$ | **Calculated Re. (%)** | **Measured Re. (%)** |
|---|---|---|---|---|
| $C_{60}$-2GUA$^+$-$\alpha$-CD | 0.477 | 0.523 | 15.3 | 36.8 |
| $C_{60}$-2GUA$^+$-$\beta$-CD | 0.516 | 0.484 | 15.8 | 33.9 |
| $C_{60}$-4OH-$\beta$-CD | 0.539 | 0.461 | 30.3 | 61.0 |

**Note**: Calcd. Residual % $= \omega_{CD}Re_{CD} + \omega_{CD}Re_{C_{60}}$; $\omega_{CD} = \frac{N_{CD}M_{CD}}{N_{CD}M_{CD}+N_{C_{60}}M_{C_{60}}}$; $\omega_{C_{60}}$=1-$\omega_{CD}$. The $\omega_{CD}$ is mass ration of CD in the corresponding complex, $Re_{CD}$ is the residual of pure CD obtained from TGA, $Re_{C_{60}}$ is the residual of pure $C_{60}$ obtained from TGA, $N_{CD}$ is the number of CD in one complex molecule, and $N_{C_{60}}$ is the number of $C_{60}$ in one complex molecule. All the detail parameter based on TGA curves are listed in Table S4.

The TGA test in the air conditions of complex $C_{60}$-2GUA$^+$-$\alpha$-CD and $C_{60}$-2GUA$^+$/$\alpha$-CD mixture was also recorded and provided in Figure S42 and S43. It is reported that CD is a polyhydroxy compound, which acts as a carbon source in the flame retardant process.[55] The same is true for $C_{60}$-4OH. $C_{60}$ itself is reported as a free radical sponge. And its own N source can be further used as flame retardant gas source.[56] Therefore, in the combustion process it can provide a certain flame retardant effectiveness. We also supplemented TGA in air condition of complex and mixture. The data mainly show two main degradation stages, and cyclodextrin self-degradation near 250 °C. By analyzing the TGA data in air, we found a significant "sustained release" phenomenon of $C_{60}$. Specifically, in complex, the decomposition phase of $C_{60}$ starts at 347 °C and ends at ~650 °C. The maximum thermal decomposition rate is less than 0.4 % °C$^{-1}$. For the blends, the pyrolysis range is ~400 °C to 650 °C, and the maximum decomposition rate is higher than 1.4 % °C$^{-1}$. This remarkable "sustained release" is helpful for $C_{60}$ to perform full as the "free radical sponge".

## 4. Methodology

**Table 3** Comparison with the conventional method for preparation $C_{60}$ (or $C_{60}$ derivatives) containing complex

| Methods | Raw materials | Solvents | Temperature (°C) | Time | Product status | Molar ratio of CD:C60 | Yield (%) | Ref. |
|---|---|---|---|---|---|---|---|---|
| Komatsu's method | $\gamma$-CD, $C_{60}$ or $C_{60}$ derivatives | - | - | >20 min | Solution | 2:1 | - | 57 32 |
| Mixed-solvent heterogeneous reaction | $\gamma$-CD, $C_{60}$ or $C_{70}$ | Water/Toluene | 118 | >30 hours | Purple crystals | 2:1 | - | 19 |
| Komatsu's method: Solid-state mechanochemical process | Functionalized $\gamma$-CD, $C_{60}$ | - | - | 20 min+20 min | Solution 70 mM | 2:1 | - | 41 |
| Komatsu's method: Solid-state mechanochemical process | 6-amino-cyclodextrin, $C_{60}$ | - | - | 20 min+20 min | - | 2:1 | - | 20 |
| Mixed-solvent heterogeneous | $\beta$-CD, pyridine | DMF/Toluene | 90 | 2.5 days | Solid | 2:1 | 69-78 | 24 |

| reaction | substituted $\beta$-CD and $C_{60}$ | | | | | | | |
|---|---|---|---|---|---|---|---|---|
| polarity induced reversible recognition | $C_{60}$ derivatives, $\alpha$-CD and $\beta$-CD | DMF+ (MeOH or $H_2O$) | R.T. | Instantaneous | Powder | 1:1 | >90 | This work |

As listed in **Table 3**, we systematically compared a series of complex preparation method and parameters. In previous reports, $C_{60}$ was included in CD by combining $C_{60}$ and CD in mixed solvent with low concentration [34], leaving for several days or weeks, or with low yield. [19] We conclude the method in this work has the significant advantages of mild condition (room temperature), high efficiency (instantaneous), high yield (> 90 %), and direct obtaining of solid products. This method can be widely and efficiently used to prepare complex, such as -OH, -COOH and other polar groups functionalized $C_{60}$. It lays a foundation for the practical application of complex. Indeed, the current preparation methods also have shortcomings. For example, access of organic solvents may be needed to remove raw materials in the process of product purification. For this point, we can use slightly excessive cyclodextrin to make fullerene derivatives fully consumed. At the same time, only appropriate amount of water is needed to remove the excess cyclodextrin in the purification stage.

Moreover, the complex reported in this study has both solvent reversibility and ionic reversibility. The structure formation rate in this work is fast. As widely reported in the articles, [58] the authors chose $\gamma$-CD as the object and used solid-state reaction. The authors selected several fullerenes for the complex preparation and 1:2 of complex was obtained. In addition, most fullerenes are modified by non-polar groups. The ratio of this work is 1:1 and we can obtain the complex highly efficiently. The current work is the expansion of previous work, and it is a novel view on the $C_{60}$ and CD based complex. As far as we know, this work reports the complex of α-CD for the first time, which expands the understanding of the complexes. As well reported, CD has the function of catalytic carbonization. [59-60] $C_{60}$ is reported to be a free radical sponge,[61] and is widely used as flame retardant.[62] Therefore, we believe that the complexes obtained by combining the $C_{60}$ and CD will also be a kind of effective flame-retardant additive.

## 5. Conclusions

In this work, we developed a simple and highly efficient method to obtain $C_{60}$-CD complex, which is termed as polarity induced reversible recognition. This method can be extended to the highly efficient combination of polar $C_{60}$ derivatives and classical cyclodextrins ($\alpha$-CD and $\beta$-CD). As typical examples, three typical complexes ($C_{60}$-2GUA$^+$-$\alpha$-CD, $C_{60}$-2GUA$^+$-$\beta$-CD and $C_{60}$-4OH-$\beta$-CD) were successfully synthesized and confirmed to be 1:1 inclusion. The geometric relationship between $C_{60}$ and CD was obtained by mathematical calculation, which had practical significance to better understand the physical structure of the complexes. The FTIR signals at near 1510 cm$^{-1}$ is detected and assigned to be the characteristic vibration signal of complex, by which the solvent reversibility of three complexes was confirmed. It was further found that after the formation of complex, the thermal stability of $C_{60}$-CD complex was significantly improved. In particular, $C_{60}$-4OH-$\beta$-CD had the highest thermal stability due to intermolecular crosslinking. It is worth noting that due to the advantages of mild conditions, high efficiency, high yield, solvent saving and formation reversibility, the new pathway laid the practical foundation in large scale application of $C_{60}$-CD complex, and the corresponding complexes have potential advanced application in the fields of flame retardant, supramolecular assembly, and molecular recognition-based sensing.

**Associated content**

Supporting Information Available: Experimental section, Figure S1-Figure S43, and Table S1-Table S4.

**Declarations of interest**

The authors declare no competing financial interest.

## Acknowledgements

This work was supported by joint Research Fund for Overseas Chinese, Hong Kong and Macao Young Scholars (51929301). The authors thank Dr. Jianfang Cui, Dr. Xiaolin Qi, Dr. Xiaomei Yang, the editor and all the reviewers for their comments on the manuscript.## References